# Power Flow Modelling of Dynamic Systems
## Introduction to Modern Teaching Tools

G.-H. Geitner[*1], G. Komurgoz[2]

Electrotechnical Institute, TU Dresden, Electrical Engineering Department, Istanbul Technical University

01062 Dresden Germany, 34469 Istanbul Turkey

[*1]Gert-Helge.Geitner@tu-dresden.de; [2]komurgoz@itu.edu.tr

*Abstract-* As tools for dynamic system modelling both conventional methods such as transfer function or state space representation and modern power flow based methods are available. The latter methods do not depend on energy domain, are able to preserve physical system structures, visualize power conversion or coupling or split, identify power losses or storage, run on conventional software and emphasize the relevance of energy as basic principle of known physical domains. Nevertheless common control structures as well as analysis and design tools may still be applied. Furthermore the generalization of power flow methods as pseudo-power flow provides with a universal tool for any dynamic modelling. The phenomenon of power flow constitutes an up to date education methodology. Thus the paper summarizes fundamentals of selected power flow oriented modelling methods, presents a Bond Graph block library for teaching power oriented modelling as compact menu-driven freeware, introduces selected examples and discusses special features.

## I. INTRODUCTION

Teaching dynamic system modelling should easily enable students to study the interactions between the power variables. From the outset they should be taught by means of tools which both clearly differ between model parts without (control) and with power flow (actuator and plant) and support an evident graphical representation. Moreover, for didactical reason it is desirable to apply methods which offer a well thought over distinction of the particular elements. Thus modern education methods should not be unique to get a dynamic model itself, but also highlight and support tasks of current interest such as power flow and energy efficiency, comparison of different system components and structures or energy distribution and recovery.

Furthermore dynamical system modelling often includes several energy domains. But the students should early get qualified to realise analogies and to overcome bounds of any specific field such as mechanical engineering or electrical engineering. In doing so one specific field should not be mapped to another specific field as known for electrical networks but teaching power flow oriented modelling should be based on a close limited and clearly laid out set of modelling elements which are independently of energy domains. Another important feature concerns advantageous definitions of very compact models in order to get detailed representations of inner configurations of the studied system for exact explanations of inner effects and power flows. Above all the reference to already taught education topics has to be ensured such as application of well-known analysis tools and control methods. The availability of a modelling method on popular software systems without any need for simulator interfaces or for introduction of specific software contributes to the fulfilment of these specifications. This explicitly includes the avoidance of iconic elements of unclear inner construction, i.e. without disclosure of equations in question.

## II. FUNDAMENTALS

Conventional modelling methods based on signal flow, i.e. transfer functions and standard block diagrams, do not consider energetic aspects, obscure the view inside the dynamical system, destroy the physical structure by changing it into a computational structure and make it difficult to compute the energy efficiency. These disadvantages vanish by use of power flow oriented methods such as Energetic Macroscopic Representation (EMR, Bouscayrol 2000) [1], Power Oriented Graph (POG, Zanasi 1991) [2], Bond Graph (BG, Paynter 1959) [3], Power Flow Diagram (Schönfeld 2004) [4] or Multipole Diagram (Mann 1975) [5]. Relationships between first four methods are very closely. These methods based on the action-reaction principle may apply same parameter definitions. For a comparison see [6].

POG and EMR explicitly present both transmission directions of the conjugated power variables of a specific connection. BG implicitly contains this information and offers very compact representations if needed and best preconditions for a universal block library. Nevertheless it has to be pointed out that lessons on the given subject definitively should include all first three methods. This may be found as well by didactic questions and different advantages. Thus the paper presents short introductions to POG, BG and EMR as well as selected modelling examples with application solutions for each method. The POG method is very suited to start teaching power flow oriented modelling on the one hand and to stimulate a discussion on the other hand, whereas BG is the most universal method and EMR provides advantages for control design particularly with regard to a common application with Causal Ordering Graph (COG) [7].

As evident from Fig. 1 power flow modelling clearly differs from signal flow modelling after having same starting point (path 1/2). Two characteristics are eye

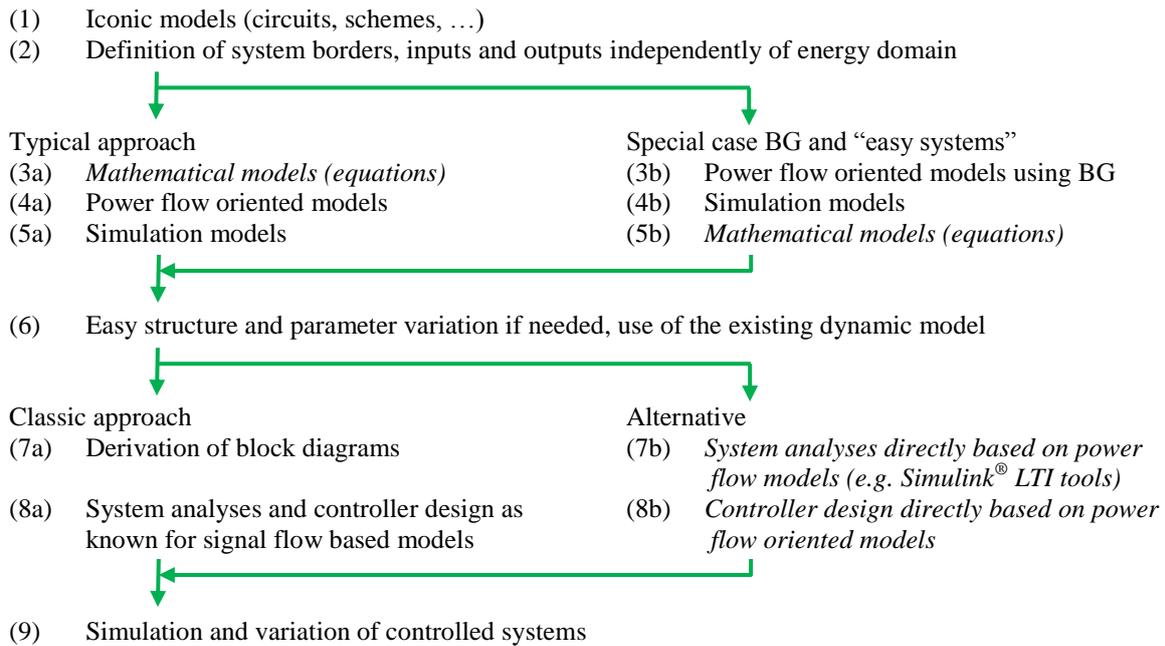

(1) Iconic models (circuits, schemes, …)
(2) Definition of system borders, inputs and outputs independently of energy domain

Typical approach
(3a) *Mathematical models (equations)*
(4a) Power flow oriented models
(5a) Simulation models

Special case BG and "easy systems"
(3b) Power flow oriented models using BG
(4b) Simulation models
(5b) *Mathematical models (equations)*

(6) Easy structure and parameter variation if needed, use of the existing dynamic model

Classic approach
(7a) Derivation of block diagrams
(8a) System analyses and controller design as known for signal flow based models

Alternative
(7b) *System analyses directly based on power flow models (e.g. Simulink® LTI tools)*
(8b) *Controller design directly based on power flow oriented models*

(9) Simulation and variation of controlled systems

Fig. 1 Typical approach using power flow modelling

catching. There is no normalization necessary which is always implied for signal flow modelling. And secondly some cases allow to model, simulate and research the system without having established any equations previously (path 3b/5b). After dynamic model study (path 6) controller design may be based both on classic (path 7a/8a) and on alternative (path 7b/8b) method.

*A. Fundamentals of BG Modelling*

Bond graph fundamentals may be outlined as following and lead to four possible connection / causality variants (Fig.2) as well as three groups of basic elements varying in the number of power ports.

*1) Connections*:

- Bonds (connections) are strictly *bidirectional*.
- *Half arrows* mark preferred directions for power transfers.
- Couples of *conjugated power variables* (e / f) are attributed to the bonds.
- Products of effort e and flow f have to result in a power value of unit Watt.
- Effort e is situated at half arrow side by definition.
- Causality strokes „|", perpendicular to one bond side, define transfer directions of flow f by definition.
- Reference direction and causality are independently of each other - compare Fig. 2.

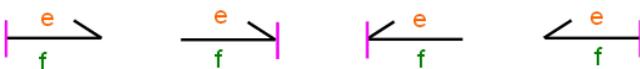

Fig. 2 Half arrow connections and causality – possible variants

*2) Basic Elements*:

- 1-Port: source / sink (S), loss element (R), energy storage (I, C);
- 2-Port: transformer, gyrator;
- multi-port: node of type 1 or 0.

Field elements for energy storage (IF, CF) and loss (RF) as well as active bond elements (AB) for measurement supplement basic element functionality. Table I gives icons including bond, parameter specification and causality options assuming integral causality to be the preferred one for storage elements. Please note causality independently parameter definitions related to the first one declared causality alternative. The two most right columns exemplify underlying internal operations for better comprehension only and summarize some hints for software implementation.

In addition to functionalities given in Table I the method defines integrals for effort and flow as generalized Momentum and Displacement and operates both with nonlinearities and initial values. Using letter "M" (modulated) as prefix for some basic elements declares the parameter to be non-constant, e.g. MR, MTF, MGY, MSE or MSF. This option needs parameter inputs via additional unidirectional powerless entrances characterised by means of normal arrows.

*3) More Hints*:

- The system representation may be done classical with scalar or optionally vectorial operations. The latter requires effort and flow vectors and parameter matrices. See also next paragraph. This option needs to take in consideration the right power for each element of the vectorial ports. Moreover 2-port elements then need to transpose in one direction.
- The method also includes structure shift at which arithmetic loops have to be considered as known.
- There are available diverse rules for configuration and reshaping of BG models.

TABLE I BOND GRAPH BASIC ELEMENTS

| BG Element; Port Type | Causality | Bond, Icon, Parameter | Resolved Presentation | Hints |
|---|---|---|---|---|
| I type energy storage; 1-Port | integral | 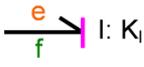 I: $K_I$ | 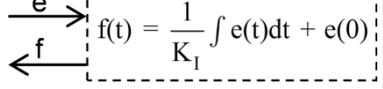 $f(t) = \frac{1}{K_I} \int e(t)dt + e(0)$ | optionally: unidirectional output ports for: power, displacement, momentum; vectorial operation |
| | differential | 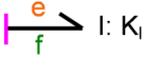 I: $K_I$ | 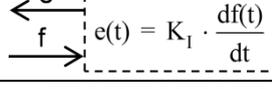 $e(t) = K_I \cdot \frac{df(t)}{dt}$ | |
| C type energy storage; 1-Port | integral | 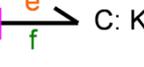 C: $K_C$ | 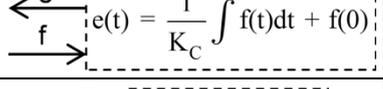 $e(t) = \frac{1}{K_C} \int f(t)dt + f(0)$ | |
| | differential | 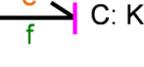 C: $K_C$ | 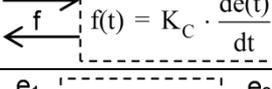 $f(t) = K_C \cdot \frac{de(t)}{dt}$ | |
| GY gyrator energy converter; 2-Port | outer | 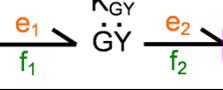 $K_{GY}$ GY | 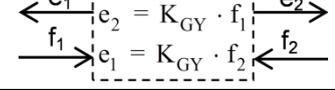 $e_2 = K_{GY} \cdot f_1$ ; $e_1 = K_{GY} \cdot f_2$ | optionally: unidirectional input port for non-linear parameters; vectorial operation including transpose operation for one direction |
| | inner | 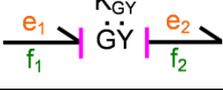 $K_{GY}$ GY | 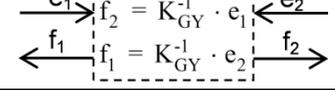 $f_2 = K_{GY}^{-1} \cdot e_1$ ; $f_1 = K_{GY}^{-1} \cdot e_2$ | |
| TF transformer energy coupler; 2-Port | left | 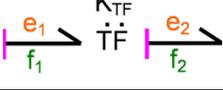 $K_{TF}$ TF | 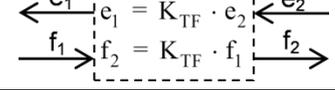 $e_1 = K_{TF} \cdot e_2$ ; $f_2 = K_{TF} \cdot f_1$ | |
| | right | 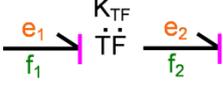 $K_{TF}$ TF | 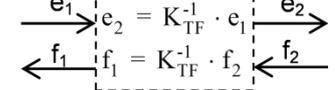 $e_2 = K_{TF}^{-1} \cdot e_1$ ; $f_1 = K_{TF}^{-1} \cdot f_2$ | |
| SE type energy source; 1-Port | E source / F sink | SE: $K_E$ 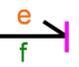 | 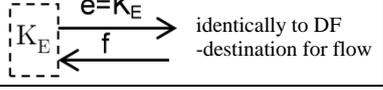 $e = K_E$, identically to DF -destination for flow | optionally: unidirectional I / O ports for control / power; vectorial operation |
| SF type energy source; 1-Port | F source / E sink | SF: $K_F$ 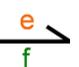 | 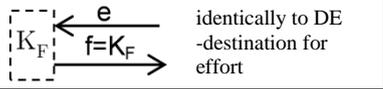 $f = K_F$, identically to DE -destination for effort | |
| R type loss element; 1-Port [inclusive power output as source thermal optionally] | Flow | 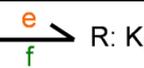 R: $K_R$ | 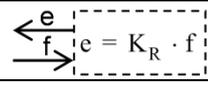 $e = K_R \cdot f$ | optionally: I and O port for parameter and power; vectorial operation |
| | Effort | 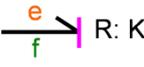 R: $K_R$ |  $f = K_R^{-1} \cdot e$ | |
| 1-node energy distributor for flow constant; lossless multi-port | at node side exactly once no causality stroke | 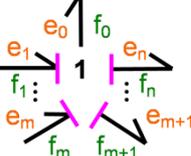 | $e_0 = + (e_1 + \ldots + e_m) - (e_{m+1} + \ldots + e_n)$ $f_0 = f_1 = \ldots = f_n$ | optionally: equation reconfiguration; vectorial operation |
| 0-node energy distributor for effort constant; lossless multi-port | at node side exactly once one causality stroke | 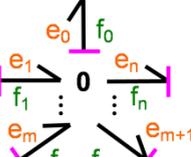 | $f_0 = + (f_1 + \ldots + f_m) - (f_{m+1} + \ldots + f_n)$ $e_0 = e_1 = \ldots = e_n$ | |
| activated bond; 1-Port | nodes power balance unchanged | 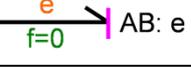 AB: e | 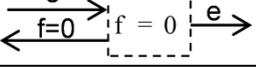 $f = 0$ | effort measurement |
| | | 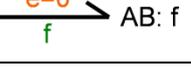 AB: f | 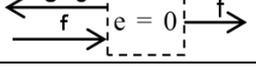 $e = 0$ | flow measurement |

Optionally bonds may be numbered only but not fit with effort and flow symbols. In fact it depends on intension and complexity. Nodes again may be additionally provided with indices in order to indicate positioning. Verbal and graphical BG design and transformation rules are given in [8] for instance.

*B. Fundamentals of POG Modelling*

The method is self-explaining and thus predestined for introduction to power flow oriented modelling. There are two types of basic elements only the elaboration block for energy storage and losses and the connection block for energy conversion. Dynamic system equations have to be reconfigured to fulfil the demands of Fig. 3. POG's are typically given in Laplace domain and show directly all mathematical operations. Unlike BG's forward and backward connections between two elements are explicitly visible. Quartered circles symbolise summing points and consequently define a maximum of 3 inputs with one output. Negative signs are marked by means of blackened corresponding input quarters.

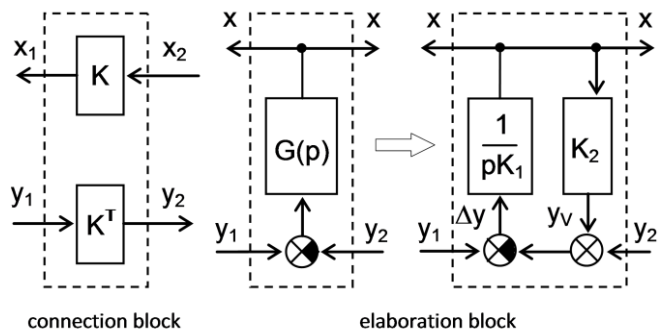

Fig. 3 POG basic elements (x, y: effort or flow; K, $K_x$: parameters)

As true for BG parameters may be of type scalar, vector or matrix as the case may be. Any product of conjugated power variables x and y has fit to a power value and for this reason power observing is possible at any point of the POG including mixing point outputs. Connection blocks directly include the mathematical transpose operation in one transmission direction and match 2-port BG elements TF and GY, summing points match BG nodes and elaboration blocks match BG loss and storage elements. The latter also have to take more sophisticated loss and storage field functionality – see paragraph IV BG block library below. Because of the POG's simplicity special definitions for sources and measurements are omitted. Connectors may improve clearness via placing parallel paths aside – cp. paragraph V example A simplification.

*C. Fundamentals of EMR Modelling*

In the same way as BG and POG the EMR approach is a graphical tool based on the action-reaction principle. EMR especially focuses the modeling on energy distribution, i.e. coupling devices as key components of the energy management in dynamic systems. This method highlights the necessity for introducing energy distribution criteria in control structures which may be obtained via step by step system inversions and decomposition into elementary subsystems. Because it leads to a macroscopic description of the whole system, other properties of the system are not pointed out.

Specific power components are represented by different elements associated to special pictograms. Historically defined geometrical icons depended on considered energy domains. Thus conversion and coupling in mechanical domain applied triangular pictograms, electrical domain applied square pictograms and circular pictograms symbolized electromechanical transformations. A next step additionally defined domain independently conversion and coupling symbols (hexagon) in order to expand application limits beyond electro-mechanical domain and finally up-to-date version only classifies mono- and multi-physical icons – see Table II. Association rules have been defined for element connections. These rules can lead to global fictive equivalent elements by a free choice of state variables due to holonomic constraints as true for BG and POG.

The EMR representation clearly shows couplings among elements and energy flows through the systems. The structure is easy to read. However, it does not show mathematical details of the model because different mathematical equations may be hidden under same icons. Thus as directly conspicuous feature results a non-representation of mixing points and signs. The method requires integral causality and subsumes energy storing and losses in one element named accumulation.

## III. SYSTEMATIZATION

Independently of any applied modelling method possible connections inside power flow models firstly may be split based on connection type scalar and vectorial. At which a further subdivision of vectorial models into systems with partially similar functionality of components and systems with in exactly the same way repeating basic segments appears usefully – see Table III. In order to complete the suggested systematization mixed systems would constitute type IV systems. Scalar type I direct modelling from physical scheme – see Fig. 1 path 3b/5b – is clearly limited to methods making use of icons, i.e. not applicable by POG, and of course requires a certain experience.

TABLE II PICTOGRAMS FOR EMR POWER ELEMENTS (definitions: [1)] historical, [2)] up-to-date 2013)

| | | energy conversion | | | | energy coupling - examples | | |
|---|---|---|---|---|---|---|---|---|
| general energy source / destination | | | | | | | | … |
| | A[1)] | electrical | electro-mechanical | mechanical | general | electrical | electro-mechanical | … |
| general accumulation and / or loss | B[2)] | mono-physical | multi-physical | - | - | mono-physical | multi-physical | - |

TABLE III TYPICAL MODEL TYPES

| Scalar | Vectorial | |
|---|---|---|
| **Type I - basic** | **Type II – similar component functionalities** | **Type III – repeating basic segments** |
| - connections of elements scalar<br>- direct modelling from physical scheme without equations optionally<br>- field elements based on matrix type parameters may connect system parts | - coordinate transform optionally<br>- folding of similar system parts enables most compact models<br>- holonomic constraints lead to equivalent subsystems | - large system modelling via segments of typically homogenous state<br>- equal parameters optionally<br>- system structure may feature an open or closed chain |
| **Examples** | | |
| - DC machine and elastic shaft<br>- solenoid | - electric three-phase machines<br>- mechanical power split solutions | - electric cable and power transformer<br>- belt conveyor |

But this possibility may cause various positive effects by concentration on the inner mode of operation, i.e. the power flow, at the outset and helps the learner to compare system features and structures at an early stage of education if taught. This perception may be obstructed via equations as known. DC machine and elastic shaft are well suited for such practice. In contrast solenoid exemplifies the need for description of mutual influences of scalar parts of a system and the use of a non-basic field element – see paragraph V.

Vectorial type II models make use of representations via components and optionally coordinate transform. Furthermore such systems possess features either of folding or of holonomic constraints or even both. Electric three-phase machine models with components stator and rotor and based on power conserving Park transform allow two-stage folding and thus both different modelling levels and very compact models if needed [9]. Same is true for planetary gears based on three connections [10]. Mechanical power split devices as systems with holonomic constraints lead to several completely equivalent model structures of same order. Moreover, equivalent dynamic system modelling again may be implemented via various solutions such as optionally use of field elements or scalar or vectorial BG elements [9, 10].

Modelling type III systems implicates two basic steps. The first step has to define a typical scalar basic segment model whose repeated usage would model the complete system. Since such large models are disadvantageously the second step involves a very compact vectorial complete model based on the first step. The mechanical Kelvin-Voigt element illustrates the idea of a basic scalar type III segment. From further studies of system types III arise subdivisions into systems with open chains, e.g. electric cable or high-frequency power transformer models, or closed chains, e.g. belt conveyor models, of basic segments.

## IV. BLOCK LIBRARY

All considered modelling approaches advantageously may apply the idea of subsystems. However, POG's do not need any definition of a block library but use standard blocks because of the simplicity [11], whereas EMR takes a collection of empty icons to be completed by the user by reason of different underlying equations [12]. Library definition makes sense for BG only [13].

### A. Simulink® add-on BG Block Library

There are a lot stand-alone and add-on software solutions to simulate BG's [8]. Indeed above summarized BG basics could provoke restrictions of the BG method to special software. But Simulink® as well established tool for diverse engineering analysis is suitable to assist students power flow oriented skills. Hence Table IV presents a clear structured add-on Simulink® freeware library [13]. Essential requirements were: a minimum number of library blocks via menu driven customization, avoidance of any editor development or compiler software and automatic realization of bidirectional connections by visibility of one direction as usually. In so doing essential equation reconfigurations, similar functionalities and possible causalities have to be identified. This leads to a combination of node types, storage elements, storage fields, sources and activated bonds. Completion by means of loss element, loss field, transformer and gyrator results in 9 basic BG library blocks only and guaranties a very well overview. Parameters may be changed during simulation if appropriate.

### B. Menu-Driven Customization

Table IV column 4 gives the possible switch-over of blocks basic functionality and meanings of defaults. Customization examples with explanations may be learned from Table IV columns 5 and 6. Block menus based on check boxes and pop-up menus permit a reasonable customization such as causality including name matching E or F, constant parameters or extra unidirectional parameter input NL, number of power ports E and F, mode of operation, output of power P and / or more outputs as well as input of initial values where applicable. This user done customization may also be locked optionally in order to avoid unintended changes. Assumed a fit parameter choice, as scalar, vector or matrix of correct size, same BG structure automatically may work scalar or vectorial. An integrated check generates warnings in case of parameter mismatch or obviously false connections between flow and effort power variables or vice versa. This flexibility primarily originates from Simulink® construction commands and tag functionality. The user is not involved as it is done automatically. Concerning graphical representations the "forward" transferred power variable always will be visualized as usual. But the "backward" one will be taken over invisible by tags.

TABLE IV SIMULINK® BOND GRAPH LIBRARY V.2.1 - BG ELEMENTS
([a] exclusive of unidirectional parameter input, [b] exclusive of unidirectional output, [c] typically ≥3, [d] variable dimension)

| Function | Library Icon | Ports | Possible Switch-Over // Default | Customization Example / Explanation |
|---|---|---|---|---|
| Source / Destination | "S" E | 1[a] | „S" == SE ↔ SF; DF; DE // constant source effort (SE) | MSE (modulated SE) as flow destination, external parameter (S), |
| Node | E1 "1" E0 F2 | multi[c] | „1" == 1-node ↔ 0-node; causality E ↔ F // 1-node: E0=E1-E2; F0=F1=F2 | 0-node distributor for effort: E1=E2=E3; F1=F2+F3 |
| Loss | F "R" FC | 1[a] | causality F ↔ E // flow causality R | effort causality, power output |
| Energy Storage | E "I" IC | 1 | „I" == I-type ↔ C-type; causality E ↔ F // integral causality, no additional output | integral C-type storage, additional momentum and power output |
| Transformer | F1 "TF" F2 LC | 2[a] | causality F ↔ E // left side (flow) causality TF | MTF (modulated TF), external parameter(NL) and right side causality |
| Gyrator | F "GY" E OC | 2[a] | causality F ↔ E // inner causality GY | MGY (modulated GY), external parameter(NL) and outer causality |
| Loss Field | F1 "RF" F2 FC | 2[d] | causality F ↔ E // flow causality R-field | MRF (modulated RF), ext. par. (NL), power output, mixed causality |
| Storage Field | F1 "CF" F2 IC | 2[d] | „CF" == C-field ↔ IF (I-field); causality mixed: E1, F1 // integral causality C-field | I-Field (IF), integral causality, power and displacement output |
| Activated Bond | F "AB" f E=0 | 1[b] | E=0 ↔ F=0 // flow measurement active bond | effort measurement |

## C. Application Hints

Non-linear parameters may be computed inside standard subsystems as usual, inputted to unidirectional BG library block inputs NL and controlled via any measured power variables as well as via any unidirectional general momentum and displacement outputs.

Any necessary model parameter itself may be provided via BG library block masks directly or automatically via special parameter definition files based on call-back functionality and file name identity plus appendage "_P". Power variable measurement exclusively has to be achieved by means of activated bonds either via another node "output" power port E / F or via incorporation of an additional "measurement" node block in order to access at the desired power value in "forward" direction. Thus a standard scope block as well can be connected to an activated bond output only because of its unidirectional operation. From the latter clearly results that interactions of BG library based models and standard blocks may only be organized via unidirectional source block inputs and unidirectional activated bond outputs. Please note, use of Simulink® LTI analysis tool input and output definitions are exceptions of this rule.

## V. EXAMPLES

Three examples hint at the immanent possibilities of the above presented tools and demonstrate substantial similarities as well as differences in graphical aspect but do not focus on application details. Schematic diagrams and associated physical equations are implemented as starting point for these power flow modeling examples to be specified as BG, POG, EMR and Simulink® BG assuming all integral initial values to be zero for simplification.

### A. Lift a Load

Figure 4 shows a basic system "lift a load". Equation system (1) assumes left shaft and rope to be non-ideal, i.e. elastically, and makes use of parameters as follows: $J_M$, $K_{FM}$ and $J_D$, $K_{FG}$ inertia / friction of motor as well as gear plus rope drum; $K_{DS}$, $K_{SS}$ and $K_{DR}$, $K_{SR}$ damping coefficient / spring rate of left shaft as well as rope; $i_G$ gear transmission ratio; $r_{DR}$ radius of drum; $m_L$ load mass and $F_g$ force of gravity. Power variable $T_M$ symbolizes the motor torque whereas symbols $\Delta T_M$, $\Delta T_{GD}$, $\Delta F$, $\Delta \omega$ and $\Delta v$ stand for dynamic torques, force and speeds.

Using Fig. 4 information Fig. 5 gives self-explaining BG, POG and EMR models. BG's utilize 0-nodes each with to simulate elastic elements. These 0-nodes plus damper and

$$\omega_M = \frac{1}{J_M} \int \Delta T_M dt \quad ; \quad \Delta T_M = T_M - T_{GM} - K_{FM}\omega_M \qquad (1a)$$

$$T_{GM} = K_{DS}\Delta\omega + K_{SS}\int \Delta\omega\, dt \quad ; \quad \Delta\omega = \omega_M - \omega_{GM} \qquad (1b)$$

$$i_G = \omega_{GM} / \omega_G = T_G / T_{GM} \qquad (1c)$$

$$\omega_G = \frac{1}{J_D}\int \Delta T_{GD} dt \quad ; \quad \Delta T_{GD} = T_G - T_D - K_{FG}\omega_G \qquad (1d)$$

$$F_R = K_{DR}\Delta v + K_{SR}\int \Delta v\, dt \quad ; \quad \Delta v = v_R - v_L \qquad (1e)$$

$$r_{RD} = \frac{T_D}{F_R} = \frac{v_R}{\omega_G} \quad ; \quad v_L = \frac{1}{m_L}\int \Delta F\, dt \quad ; \quad \Delta F = F_R - F_g \qquad (1f)$$

$\omega_M$ — angular speed at motor side  
$\omega_{GM}, \omega_G$ — angular speeds at both gear sides  
$v_R$ — rope circumferential speed on drum  
$v_L$ — rope speed at load side  

Fig. 4 Schematic representation and equations of a basic system "lift a load" (elastic elements red marked)

Fig. 5 Power Flow oriented models "Lift a Load": (a) BG, (b) POG, (c) EMR and (d) Simulink® BG

spring related to have to be removed if shaft and / or rope are taken for ideal. If appropriate then parameters will have to be subsumed in order to avoid irrelevant derivative causalities. BG Fig. 5a models the system featuring three independent movements ($\omega_M$, $\omega_G$, $v_L$) and therefore includes three associated energy storages of integral causality - exclusive of spring / damper models. Please note the strong structure analogy of Fig. 5a and 5d.

The presupposition of ideal elements would lead to one independent movement only and thus inertias and masses would have to be subsumed to one fictive parameter. This statement is universally valid and therefore it has to be implemented over any conversion elements of type TF or GY likewise. The position of such exclusive energy storage is optional in principle. It may be modeled as total inertia or total mass.

POG Fig. 5b ignores both dampers ($K_{DS}$, $K_{DR}$) in order to get one POG-path only. If damping effects shall be taken into account then connectors have to model this power split via parallel POG paths. Since EMR accumulation icon includes losses this easy application example doesn't demands any coupling icon for EMR Fig. 5c but produces an easy chain without branches although models all features contrary to POG Fig. 5b. EMR conversion parameters $i_{GT}$ and $r_{DR}$ are given ready for use, same as true for POG and in contrast to use of flow related BG definitions. For simplification BG's ignore any measurements of power variables and all models assume the motor as controlled torque source only. Power flow based motor models may be learned from [14] for instance.

*B. Solenoid*

An elementary solenoid system is given in Fig. 6 based on equation system (2). The non-linear dependence of inductance L(x) on armature position x is known (2e) and thus matrix $\underline{M}$ may describe mutual influence of magnetic and mechanical domain (2d). Parameters are defined as follows: n number of coil turns, R ohmic coil resistance, m armature mass, $K_{fric}$ translational friction coefficient, A cross section of the limb, $l_m$ medial length of a magnetic flux field line in iron, $\mu_0$ magnetic field coefficient, $\mu_r$ permeability of iron, $x_0$ initial position and $F_g$ force of gravity. Power variables u and i symbolize applied coil voltage and the current, $\Theta$ and $\Phi$ stand for magnetic voltage and flux whereas $F_M$, $F_{fric}$ as well as v describe magnetic and friction force plus armature speed. Again $\Delta$ symbols meet dynamic values for force and voltage. This example possesses following specific feature. Since magnetic domain shall be modeled explicitly ohmic resistance and inductivity will be modeled at separated places.

Measured current i and position x control the modulation of matrix $\underline{M}$. Hence modulated BG field of type MCF serves as a basis of non-linear BG model Fig. 7a. Blue markings graphically hint at that position x may be received from an appropriate I-type storage as a displacement. Measurement of current i via activated bond is omitted for simplification. The left-sided 1-node represents the losses of the electric circuit whereas the right-sided 1-node models the balance of forces. Simplified explicit magnetic domain modeling is done by means of a GY element applying number of turns as parameter. A 0-node, senselessly at first sight, hints at Simulink® BG Fig. 7d which doesn't allow connections of non-power distribution elements without nodes [13]. Same issue represents POG Fig. 7b but shows equations directly. For this it is necessary to sum up correctly signed three times effort quantities and once flow quantities. In this case connection blocks with parameters $\underline{A}$ and $\underline{B}$ realize vectorial composition and decomposition only. BG field elements imply this functionality anyway. EMR again subsumes blocks 3 till 5 of the POG model and applies the general conversion icon. Just as per BG there are no special elements visible for constituting vectors.

*C. Filter and Chopper*

A basic RLC circuit is shown in Fig. 8. It may model any filter or an energy link for power electronics (3a/c). Via supplying a chopper (3d) it serves as a variable energy source for DC machines. Parameters are defined as follows: $C_F$ capacitance, $L_f$ inductance, $R_F$ ohmic resistance of real inductance $L_F$ and $m_{Ch}$ chopper control. There are two possibilities to model a chopper control via parameter $m_{Ch}$.

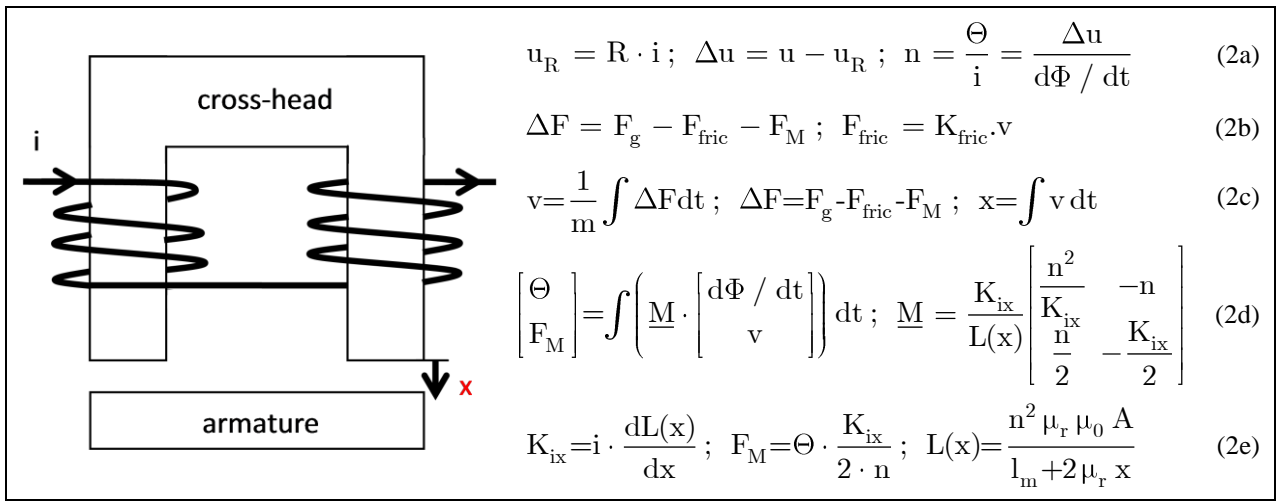

$$u_R = R \cdot i \ ; \ \Delta u = u - u_R \ ; \ n = \frac{\Theta}{i} = \frac{\Delta u}{d\Phi / dt} \quad (2a)$$

$$\Delta F = F_g - F_{fric} - F_M \ ; \ F_{fric} = K_{fric} \cdot v \quad (2b)$$

$$v = \frac{1}{m} \int \Delta F \, dt \ ; \ \Delta F = F_g - F_{fric} - F_M \ ; \ x = \int v \, dt \quad (2c)$$

$$\begin{bmatrix} \Theta \\ F_M \end{bmatrix} = \int \left( \underline{M} \cdot \begin{bmatrix} d\Phi / dt \\ v \end{bmatrix} \right) dt \ ; \ \underline{M} = \frac{K_{ix}}{L(x)} \begin{bmatrix} \frac{n^2}{K_{ix}} & -n \\ \frac{n}{2} & -\frac{K_{ix}}{2} \end{bmatrix} \quad (2d)$$

$$K_{ix} = i \cdot \frac{dL(x)}{dx} \ ; \ F_M = \Theta \cdot \frac{K_{ix}}{2 \cdot n} \ ; \ L(x) = \frac{n^2 \mu_r \mu_0 A}{l_m + 2\mu_r x} \quad (2e)$$

Fig. 6 Schematic representation and equations of a basic system "solenoid" (position red marked)

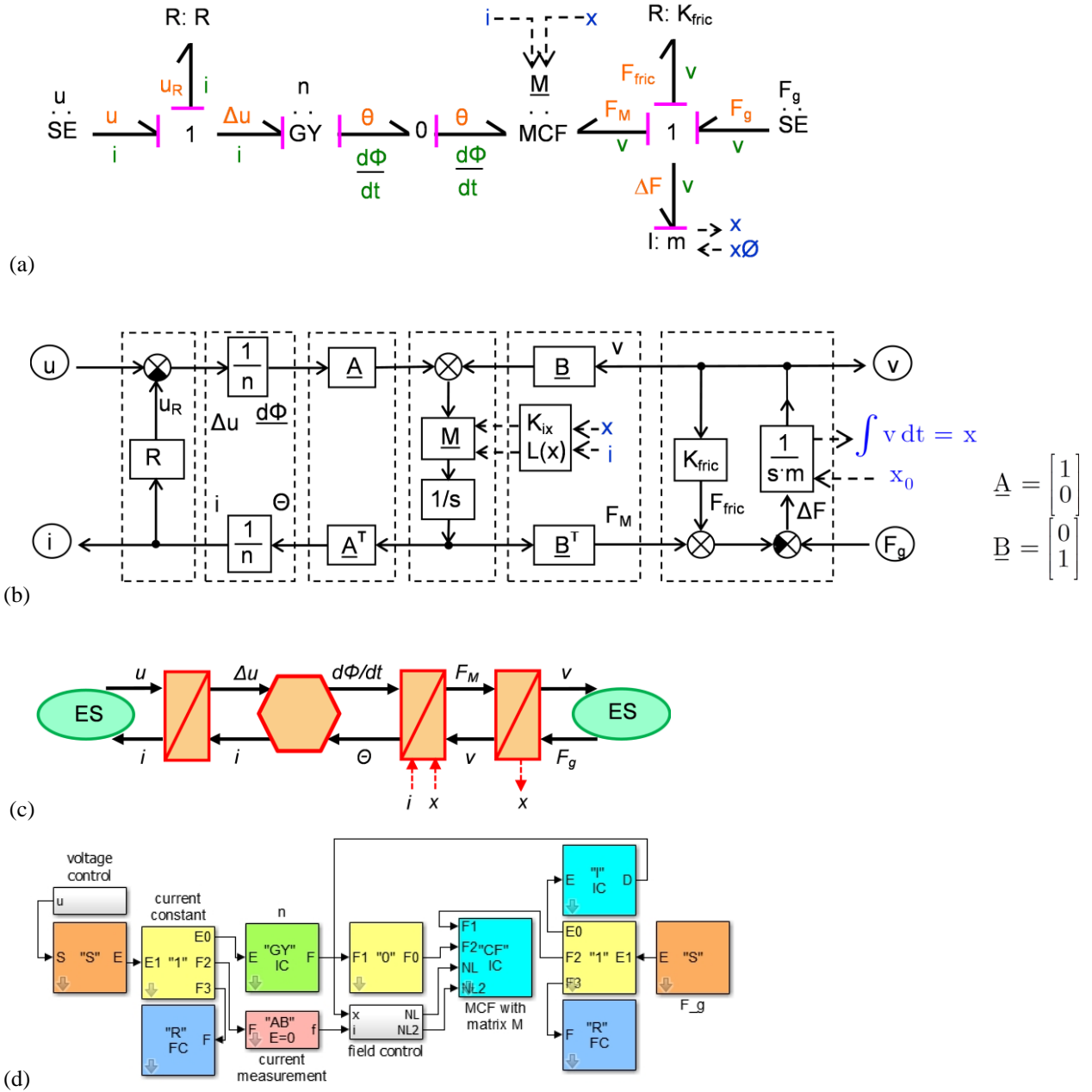

Fig. 7 Power Flow oriented models "Solenoid": (a) BG, (b) POG, (c) EMR and (d) Simulink® BG

Either high frequently pulse width modulation control will be simulated indeed or mean values for chopper output voltage will be used. Known disadvantages of stiff differential equations arise from the former. The latter enables enough accurate results in many cases if relatively large dynamic systems are modeled. For Simulink® based chopper simulation see [13].

Left side voltage mesh equation describes connection of filter input voltage $u_{in}$, chopper input voltage $u_{Ch}$, resistive voltage drop $u_R$ and inductive voltage drop $\Delta u$ as well as filter current $i_{in}$. Equations (3d) link together the input and output values of the chopper assuming an ideal chopper device (IC).

For this easy system procedure path 3b/5b in Fig. 1 also directly results in BG Fig. 9a without write down any equation. Equations (3d) transform power variables of same type using a variable parameter. That means modulated transformer (MTF) based BG representation. Analogies between Kirchhoff's voltage law and 1-nodes as well as Kirchhoff's current law and 0-nodes are obviously. Consequently POG model Fig. 9b features two elaboration blocks, one without loss part, and one connection block. Corresponding EMR model is given in Fig 9c and realizes filter output voltage and chopper output voltage connection by electrical conversion element controlled via $m_{ch}$. EMR method also may subsume functionality of both POG elaboration blocks and thus Fig. 9c may be more simplified if favoured, but then the clear graphical statement that one power variable type depends on a difference of the other type and simultaneously defines "inputs" for antecessor and successor modeling element would be seriously affected. Generally load current $i_{out}$ is taken for granted and thus right-hand side sources have to operate as effort destination

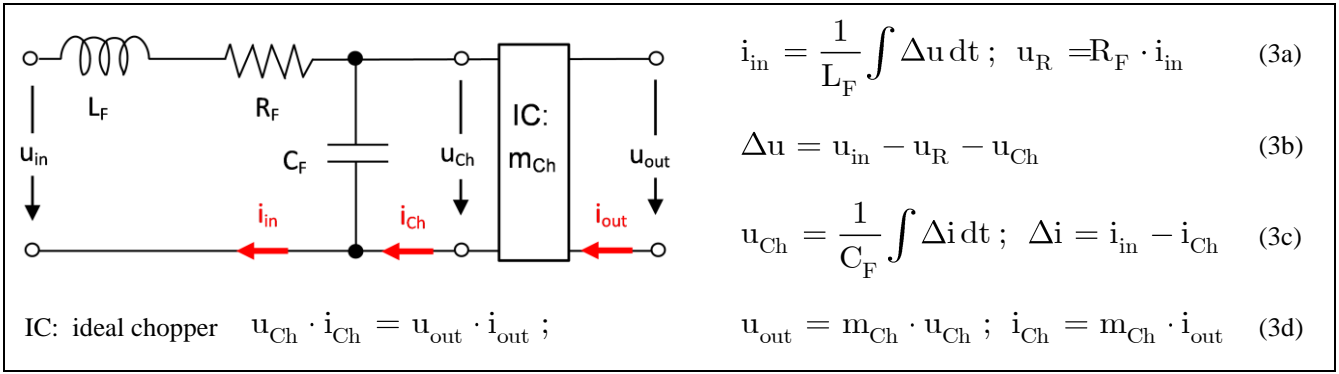

Fig. 8 Schematic representation and equations of a basic system "filter and chopper" (currents red marked)

$$i_{in} = \frac{1}{L_F} \int \Delta u \, dt \, ; \quad u_R = R_F \cdot i_{in} \quad (3a)$$

$$\Delta u = u_{in} - u_R - u_{Ch} \quad (3b)$$

$$u_{Ch} = \frac{1}{C_F} \int \Delta i \, dt \, ; \quad \Delta i = i_{in} - i_{Ch} \quad (3c)$$

IC: ideal chopper $\quad u_{Ch} \cdot i_{Ch} = u_{out} \cdot i_{out} \, ; \quad u_{out} = m_{Ch} \cdot u_{Ch} \, ; \quad i_{Ch} = m_{Ch} \cdot i_{out} \quad (3d)$

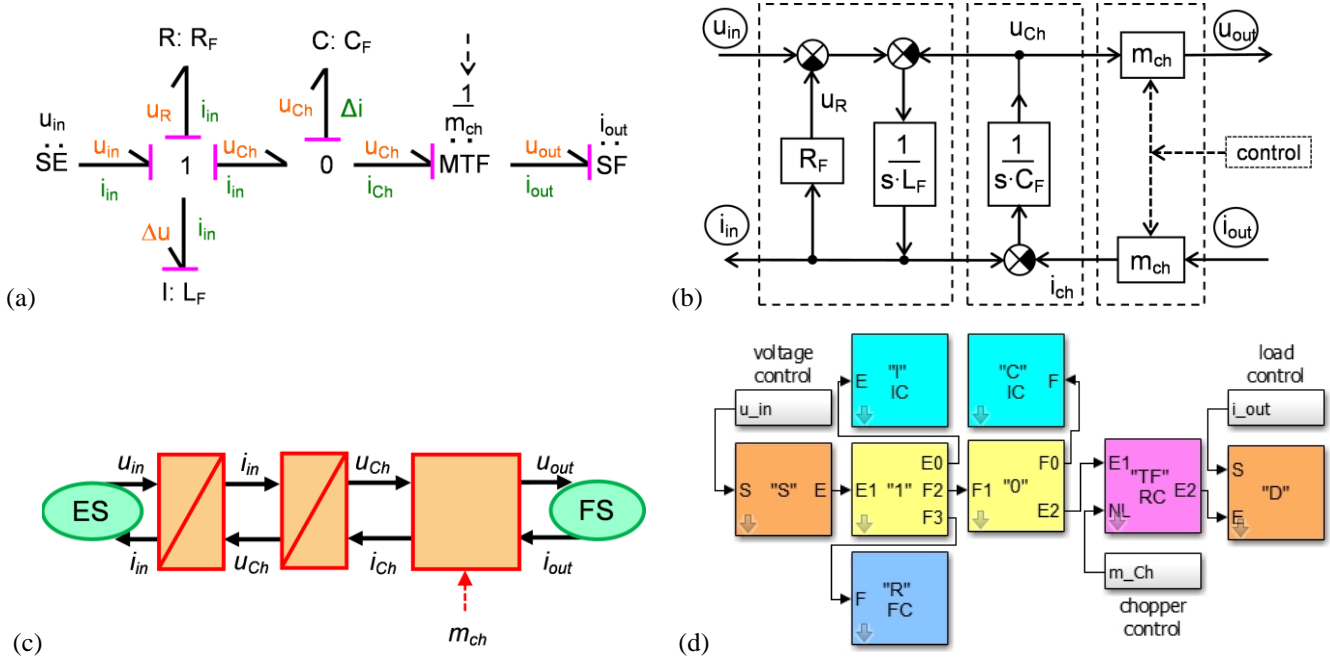

Fig. 9 Power Flow oriented models "filter and chopper": (a) BG, (b) POG, (c) EMR and (d) Simulink® BG

unlike for examples 1 and 2. All models permit a very easy addition of a possible capacitance loss resistance without structure changes or voltage and load control as adumbrated in Fig. 9d.

## VI. SPECIAL FEATURES

Although power flow oriented modelling may be applied to any energy domain and analogously to non-technical areas too some special cases have to be pointed out in order to hint at the immanent potential of such tools. The selection of course is a subjective one but shall inspire students and instructors to deal with this trendsetting modelling approach.

- The method explicitly leads the user to the principle of power consistency. This may be easy understood for research into systems input / output power. But deepens the understanding of 3-phase systems by use of power conserving Park or Clarke transform for electric 3-phase machines or decomposition into common or differential mode quantities for instance [15, 16]. Otherwise modelling would fail.

- A specified modelling via introduction of another energy domain may not only include a structure upgrading but also cause changes of storage and conversion types. This applies to the magnetic domain for electric machine models for one [17].
- Some well-known, for a long time used modelling and definitions may historically result from analogies but do not agree with power flow oriented modelling, such as true for thermal or magnetic domain. This fact facilitates more studies and examination of pseudo power flow modelling idea [3]. Otherwise it has to be stated that pseudo-BG are no different in principle and practice from regular BG [18].
- Partial systems may be modelled via vectorial power variables, fields or scalar models based on absolutely equivalent equations. This method gives a good reason to deal with equivalent conversions [9].
- The approach is open and may tolerate advancement for new challenges such as suggestions proposed for generalizing bond connections via paired information variables [19].

- State space representation may be easy extracted including time-varying state space transformation and parameter definition [20].
- Although Bond Graphs and Linear Graphs both are multi-disciplinary in principle the authors clearly prefer Bond Graphs from a pedagogical point of view and thus support the detectable asymmetry between the two methods [18].

Since modelling doesn't end in itself this approach also significantly enhances possibilities to study the systems features. Following list attests these theses. Whereat features 3 and 4 refer to Simulink® simulations but analogously are true for similar software:

- energy efficiency computation obviously easy to handle based on the modelling approach itself [14];
- all power variables and their integrals easy accessible due to a fit model structure – cp. Paragraph IIA;
- linear time invariant (LTI) analyses tools still direct disposable, e.g. [21];
- simple connection of bidirectional plant models with unidirectional control structures, e.g. [21];
- new controller generation algorithms directly based on power flow models, e.g. [7];
- direct z-transformation based digital controller design avoiding any approximations [22];
- model reduction directly based on energy flow instead of transfer functions or state space representation [23, 24];
- direct power flow modelling based topology and parameter system optimization [25].

LTI tools include standard preparations for usual controller design of conventional cascade or state control structures such as Bode diagram, Nyquist diagram, pole zero map or automatic state space and transfer function generation. Power flow methods may be easy transformed into each other if the focus of the research interest changes. Even for manual generation of usual transfer functions there are convenient rules.

VII. CONCLUSIONS

Power flow oriented modelling efficiently promotes students skills on dynamic systems. The students get a view inside the physical structure of the system, deepen knowledge about conjugated power variable pairs and turn their focus to physical background. Common simulation software is still sufficiently. Available methods are related to each other, but different in focus. POG is the best choice for beginners and shows equations immediately. BG uses icons, but equations belonging to are definitely fixed and the method may result in very compact models. EMR again uses icons likewise, but respective equations depend on applications. Generally power flow oriented research and education approaches enable quick results regarding system structures and features. Typical fields of application are automotive systems in particular and mechatronic systems in general.

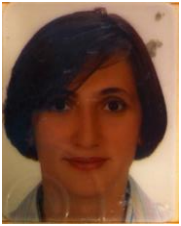

**G. Kömürgöz** completed her BS (1991), MS (1995), and PhD (2001) all in electrical engineering at the Istanbul Technical University, Turkey. She is currently an associate professor of electrical engineering. She is interested in Heat Transfer, Numerical Methods, Design of Transformer and Electrical Machines.

.

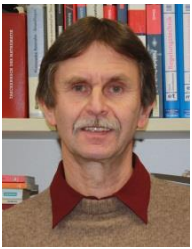

**G.-H. Geitner** completed his BS (1976), MS (1981), and PhD (1991) all in electrical engineering at the Dresden Technical University, Germany. He is currently a Privatdozent of electrical engineering. He is interested in Digital Amplitude Optimum and Bond Graph Application, Design of Digital Controllers for Electrical Drives and Power Flow Modeling methods